\documentstyle[preprint,aps,epsfig,floats]{revtex}

\newcommand{\ac}{{\centerline{\bf Acknowledgments}}}
\draft
\tightenlines
\begin{document}
\date{January 5, 1999}
\preprint{\vbox{\hbox{RUB-TPII-03/98}\hbox{PNU-NTG-02/98}}}
\title{Pion form factors with improved infrared factorization\\}
\author{N. G. Stefanis\thanks{Email:
        stefanis@hadron.tp2.ruhr-uni-bochum.de}
        and
        W. Schroers\thanks{also Fachbereich Physik,
        Universit\"at Wuppertal,
        D-42097 Wuppertal, Germany}
        \thanks{Email: wolfram@theorie.physik.uni-wuppertal.de}
        }
\address{Institut f\"ur Theoretische Physik II,    \\
         Ruhr-Universit\"at Bochum,                \\
         D-44780 Bochum, Germany                   \\
         }
\author{H.-Ch. Kim\thanks{Email:
        hchkim@hyowon.pusan.ac.kr}
        }
\address{Department of Physics,              \\
         Pusan National University,          \\
         Pusan 609-735, Republic of Korea    \\
         }
\maketitle
\begin{abstract}
We calculate electromagnetic pion form factors with an analytic
model for $\alpha _{\rm s}(Q^2)$ which is infrared (IR) finite without
invoking a ``freezing'' hypothesis.
We show that for the asymptotic pion distribution amplitude,
$F_{\pi ^{0}\gamma ^{*}\gamma}$ agrees well with the data, whereas
the IR-enhanced hard contribution to $F_{\pi}$ and the soft
(nonfactorizing) part can jointly account for the data.
\end{abstract}
\newpage   
\input amssym.def
\input amssym.tex
The issue of computing exclusive processes, like electromagnetic form
factors, within QCD is of fundamental interest because it reveals the
basic structure of hadrons. But in contrast to inclusive processes,
there is much uncertainty about the applicability of perturbative QCD
at laboratory momenta. It seems that the more interpretations this
subject inspires and grounds the more unsettled it is.

This paper attempts to bring together some crucial results about pion
form factors and combine them with novel theoretical developments
concerning the infrared (IR) regime of QCD, in order to benchmark the
status of our current understanding. Such an approach appears attractive
since the incorporation of nonperturbative power corrections in the
perturbative domain may improve both the IR insensitivity of exclusive
observables and the self-consistency of calculations entrusted to
perturbative QCD
\cite{SS97,KP96,DMW96,AA97,Gru97,AZ97,Web97,GKG98,MSS98,Shi98}.

To this end, the conventional representation of the running strong
coupling ``constant'' is given up in favor of an analytic model,
recently proposed by Shirkov and Solovtsov \cite{SS97}, which
incorporates a single power correction to remove the Landau singularity.
Bearing in mind that the definition of $\alpha _{s}$ beyond two loops
cannot be uniquely fixed, one may regard the ambiguity in the IR
modification of the running coupling as resembling the freedom of
adopting a particular non-IR-finite renormalization scheme \cite{SSK98}.

The Shirkov-Solvtsov model employs Lehmann analyticity to bridge the
regions of small and large momenta by changing the one-loop effective
coupling to read
\begin{equation}
  \bar{\alpha}_{s}^{(1)}(Q^{2})
=
  \frac{4\pi}{\beta _{0}}
  \left[
          \frac{1}{\ln \left( Q^{2}/\Lambda ^{2} \right)}
        + \frac{\Lambda ^{2}}{\Lambda ^{2} - Q^{2}}
  \right] \; ,
\label{eq:alpha_an}
\end{equation}
where $\Lambda \equiv \Lambda _{\rm QCD}$ is the QCD scale parameter.
This approach provides a nonperturbative regularization at low momenta
and leads to a universal value of the coupling constant at zero momentum
$
   \bar {\alpha}_{s}^{(1)}(Q^{2}=0)
 = 4 \pi /\beta _{0}
\simeq
   1.396
$
(for three flavors), defined only by group constants, i.e., avoiding the
introduction of external parameters, like an (effective) gluon mass to
``freeze'' the coupling at low momentum scales.

Note that this limiting value (i) does not depend on the scale parameter
$\Lambda$ -- this being a consequence of renormalization group
invariance -- and (ii) extends to the two-loop order, and beyond, i.e.,
$
 \bar {\alpha}_{s}^{(2)}(Q^{2}=0)
=
 \bar {\alpha}_{s}^{(1)}(Q^{2}=0)
\equiv
 \bar {\alpha}_{s}(Q^{2}=0)
$.
(In the following the bar is dropped.) Hence, in contrast to standard
perturbation theory, the IR limit of the coupling constant is stable,
i.e., does not depend on higher-order corrections, and is therefore
universal. As a result, the running coupling also shows IR stability.
This is tightly connected to the nonperturbative contribution
$\propto \exp (-4\pi /\beta _{0})$ which ensures analytic behavior in
the IR domain by eliminating the ghost pole at
$Q^{2} = \Lambda _{\rm QCD}^{2}$.

At very low momentum values, say, below 1~GeV, $\Lambda _{\rm QCD}$
in this model deviates from that used in minimal subtraction schemes.
However, since we are primarily interested in a region of momenta which
is much larger than this scale, the role of this renormalization-scheme
dependence is only marginal. In our investigation we use
$
 \Lambda _{\rm QCD}^{{\rm an} (3)}
=
 242~{\rm MeV}
$
which corresponds to
$
 \Lambda _{\rm QCD}^{{\overline{\rm MS}} (3)}
=
 200~{\rm MeV}
$.

This analytic model for the strong running coupling is very suitable
for calculations of exclusive amplitudes, mainly for two reasons:
Firstly, it ensures IR safety of the factorized short-distance part
without invoking the additional assumption of saturation of color forces
by using a gluon mass -- extensively used up to now in form-factor
calculations (see, for example, \cite{BJPR98}).
Furthermore, the Sudakov form factor \cite{BS89} does not have to serve
as an IR protector against $\alpha _{\rm s}$ singularities. Hence, the
extra constraint of using the maximum between the longitudinal and the
transverse scale, as argument of $\alpha _{\rm s}$, proposed in
\cite{LS92}, becomes superfluous.
This is a serious advantage relative to previous analyses because now
one is able to choose the unphysical constants \cite{CS81}, which
parametrize different factorization and renormalization schemes, in
such a way as to optimize calculated observables (for a more detailed
discussion of this point, we refer to \cite{SSK98}). Second, and more
important, the non-logarithmic term in Eq.~(\ref{eq:alpha_an}) enters
all anomalous dimensions, viz. the cusp anomalous dimension, which gives
rise to the Sudakov form factor \cite{KR87,Kor89,Col89,KR92,GKKS97}, as
well as the quark anomalous dimension which governs evolution. As a
result, the suppression due to transverse momenta, intrinsic \cite{JK93}
and those generated by radiative corrections \cite{LS92}, is
counteracted, and hence there is no reduction of the form-factor
magnitude.

We are going to show in this work that the enhancement effect
originating from the power correction to the running coupling is enough
for the asymptotic pion distribution amplitude to contribute (at
leading order) to the spacelike electromagnetic form factor of the pion,
$F_{\pi}(Q^{2})$, a hard part that can account for almost half of the
form-factor magnitude relative to the existing data \cite{Bro73,Beb76}.

On the other hand, the transition form factor,
$F_{\pi ^{0}\gamma ^{*}\gamma} (Q^{2})$, is only slightly changed,
as compared to the result given in \cite{JKR96}, and matches the
recent
high-precision CLEO data \cite{CLEO98} as good as the dipole fit. Also
the older CELLO data \cite{CELLO91} at lower $Q^{2}$ are well
reproduced (see below).

In both cases, no adjustment of the theoretical predictions to the
experimental data is involved.

Therefore, there appears to be no need to reanimate
endpoint-concentrated pion distribution amplitudes, proposed by Chernyak
and Zhitnitsky (CZ) \cite{CZ84}, in order to make contact with the
experimental data -- as recently attempted in \cite{TL98}.
We are less enthusiastic about using such distribution amplitudes
because of the following serious disadvantages: \\
(i) It has been recently shown \cite{JKR96,Ong95} that distribution
    amplitudes of the CZ-type lead to a $\pi\gamma$ transition form
    factor which significantly overestimates the CLEO data just
    mentioned. Our reasons for skepticism parallel the arguments given
    in \cite{JKR96} and will not be repeated here. The excellent
    agreement between theory (QCD) and measurement for this process,
    already at leading order, when the asymptotic pion wave function is
    used, cannot be overemhasized.
    Note in this context that the calculation of Cao {\it et al.}
    \cite{CHM96}, which predicts for the CZ wave function smaller values
    of $F_{\pi ^{0}\gamma ^{*}\gamma}(Q^{2})$ than the data, uses for
    modeling the $k_{\perp}$ distribution in the pion an ansatz that
    strongly suppresses the endpoint region. Hence, in effect, their
    wave function, though claimed to be the CZ one, excludes this region
    and yields therefore a result even smaller than the one predicted by
    the asymptotic distribution amplitude in the range of $Q^{2}$ where
    there are data. Independently, investigations \cite{RR96,MR97} based
    on QCD sum rules come to a comparably good description of
    $F_{\pi ^{0}\gamma ^{*}\gamma}(Q^{2})$ at not too low $Q^{2}$ on
    the basis of local duality without presuming the asymptotic form of
    the pion distribution amplitude, but favoring again a shape close to
    that.

(ii) Distribution amplitudes of the CZ-type yield a direct-overlap,
     i.e., soft contribution, to $F_{\pi}(Q^{2})$ which turns out to be
     of the same large order of magnitude \cite{IL84,Rad90,BH94,JKR96}
     as that resulting from the convolution with the hard-scattering
     amplitude \cite{LB80,ER80,CZ84}, or is even larger, at currently
     probed $Q^{2}$ values. Inclusion of this contribution into the pion
     form factor leads eventually to a total result which overestimates
     the existing data considerably -- even allowing for some double
     counting of hard and soft transverse momenta near the transition
     region. Though the present quality of the high-$Q^{2}$ data
     \cite{Bro73,Beb76} on the spacelike pion form factor is quite poor,
     the trend seems to be indicative.

(iii) The underlying QCD sum rules analysis of \cite{CZ84} suffers
      with respect to stability -- as outlined by Radyushkin
      \cite{Rad90}. As a result, the duality interval increases with
      moment order $N$, meaning that the (nonperturbative) condensate
      contributions grow with order relative to the perturbative term.
      A direct consequence of this is that the moments of the pion
      distribution amplitude for $N=2,4$, extracted from these sum
      rules, are artificially enhanced. Such large moment values can
      only be reproduced by a double-humped endpoint-concentrated
      distribution amplitude and correspond to the basic assumption that
      vacuum field fluctuations have infinite size, or equivalently that
      vacuum quarks have exactly zero virtuality \cite{Rad90}.

(iv) The characteristic humps in the endpoint regions ($x=0,1$) are
     not generic, but merely the result of truncating the eigenfunctions
     expansion of Gegenbauer polynomials at polynomial order two while
     keeping the normalization fixed to unity. Including higher and
     higher order polynomials, the humps become less and less prominent
     and the central region ($x=1/2$) gets enhanced. To this point, we
     mention that an independent QCD sum rules analysis \cite{BF89}
     gives the constraint
     $
      \phi _{\pi}(x=1/2)
     =
      1.2\pm 0.3
     $,
     which is close to the value
     $
      \phi _{\rm as}(x=1/2)
     =
      3/2
     $,
     and definitely violated by the CZ amplitude.

Physically, the source of the endpoint enhancement of CZ-type
distribution amplitudes can be understood as follows.
If the vacuum quark virtuality is zero, an infinite number of such
quanta can migrate from the vacuum to the pion state at zero energy
cost.
This happens exactly in the kinematic region $x=0$ or $\bar{x}\equiv
1-x=0$ and leads to a strong enhancement of that region at the expense
of depleting the amplitude for finding configurations in which the
quark and the antiquark, or more precisely, the struck and the spectator
partons, share almost equal momentum fractions around $x=1/2$.
In the pion, configurations close to the kinematic boundary contain one
leading parton, which picks up almost all of the injected momentum, and
an infinite number of wee partons $\sim 1/x$ with no definite
transverse positions relative to the electromagnetic probe, which
constitute a soft cloud. In this regime, gluons have very small
virtualities and therefore it is inconsistent to assume hard-gluon
rescattering, i.e., the factorization of a short-distance part in the
exclusive amplitude becomes invalid. This region of momenta has to be
treated separately on the basis of the Feynman mechanism just described,
but a theoretical approach from first principles, though of paramount
importance, is still lacking.
On the other hand, if the vacuum virtuality is sizeable, say, of
the order of $\Lambda _{\rm QCD}$ or even larger \cite{DEM97}, then an
energy gap might exist that prohibits the diffusion of vacuum quarks
into the pion state, and hence Feynman-type configurations are insulated
from those for which hard-gluon exchange applies. This gap may be the
result of nonlocal condensates \cite{Rad90,MR86,BM95}, which have a
finite fluctuation size, or alternatively being induced in the form of
an effective quark mass acquired through the interaction with an
instanton background \cite{Dor96,PPRWG98}.
But the general result is the same: the shape of the pion distribution
amplitude gets strongly enhanced in the central region and resembles
closely the asymptotic one.

In view of these drawbacks, a potentially good agreement between
theoretical estimates employing CZ-type distribution amplitudes for the
pion -- as recently reported in \cite{TL98} -- and experimental
measurements is entirely circumstantial.

In the present work, we are going to show that taking together the soft
form-factor contribution due to the overlap of the initial and final
pion wave functions \cite{JKR96}, and the hard, i.e., factorizing part,
we can obtain a result in qualitative agreement with the existing data
and complying with the power counting rules.
Actually, including also NLO contributions to the hard-scattering
amplitude (see, \cite{MNP98} and previous references therein),
$F_{\pi}(Q^{2})$ gets additionally enhanced to account for approximately
half of the form-factor magnitude \cite{SSK98}, modulo the large
uncertainties of the existing data.

However, we cannot and do not exclude that the true pion distribution
amplitude may deviate from the asymptotic one, but this deviation should
be within the margins allowed by the experimental data for the
$\pi\gamma$ transition form factor. Hence the true pion distribution
amplitude may well be a hybrid of the type
$
 \Phi _{\rm true}
 =
  90\% \Phi _{\rm as} + 9\% \Phi _{\rm CZ} + 1\% C_{4}^{3/2}
$.
This mixing ensures a broader shape of the pion distribution amplitude,
with the fourth-order, ``Mexican hat''-like, Gegenbauer polynomial,
being included in order to cancel the dip at $x=1/2$. The shapes derived
from instanton-based approaches \cite{Dor96,PPRWG98} are of this type.
For such distribution amplitudes, evolution already at LO must be taken
into account that tends to reduce the importance of the endpoint region
leading to a decrease of the magnitude of the form factors towards the
data \cite{JKR96,MNP98}. On the other hand, for the asymptotic solution,
evolution enters only at NLO and is a tiny effect which is ignored in
the present exploratory investigation.

The starting point of our analysis is the expression for the pion form
factor in the transverse configuration space after employing
factorization to separate a short-distance, i.e., hard-scattering part
(where the the terminology of \cite{CS81,BS89,LS92} is adopted):
\begin{eqnarray}
  F_{\pi}\left( Q^{2} \right)
= &&
  \int_{0}^{1} {\rm d}x {\rm d}y \int_{-\infty}^{\infty}
               \frac{{\rm d}{}^{2}b}{(4\pi )^{2}} \,
               {\cal P}_{\pi}^{\rm out}
               \left( y, b, P^{\prime}, \mu _{\rm F}, \mu _{\rm R}
               \right)
  T_{\rm H}\left(  x, y, b, Q/\mu _{\rm R}, Q/\mu _{\rm F} \right)
\nonumber \\
&&
\times \; {\cal P}_{\pi}^{\rm in}
          \left( x, b, P, \mu _{\rm F}, \mu _{\rm R}
          \right) \; ,
\label{eq:piffbspace}
\end{eqnarray}
Here $P^{+}=Q/\sqrt{2}=P^{-\prime}$, $Q^{2}=-(P^{\prime}-P)^{2}$,
and $\mu _{\rm R}=C_{2}\xi Q$ and $\mu _{\rm F}=C_{1}/b$ are,
respectively, the renormalization and factorization scales, with
$C_{1}, C_{2}\sqrt{2}=C_{2}^{CS}$ \cite{CS81} ($\xi = x, \bar{x}, y,
\bar{y}$), the constants $C_{1}$, $C_{2}$ being integration constants
of order unity, so that (uncalculated) higher-order corrections are
small \cite{CS81,Col89}. Finally, $b$ is the variable conjugate to the
transverse gluon momentum, and denotes the transverse distance between
quark and antiquark.

The hard-scattering amplitude $T_{\rm H}$ is the amplitude for a quark
and an antiquark to scatter collinearly via a hard-gluon exchange with
wavelengths limited by $b$, and is given in leading order by
\begin{equation}
 T_{\rm H}\left( x, y, b, Q/\mu _{\rm R}, bQ \right)
=
 8 C_{\rm F} \alpha _{\rm s}^{{\rm an}}(\mu _{\rm R}^{2})
 K_{0} \left( \sqrt{xy}\, bQ \right) \; .
\label{eq:hardamplbspace}
\end{equation}
In Eq. (\ref{eq:piffbspace}), ${\cal P}_{\pi}$ describes the valence
${\rm q}\bar{{\rm q}}$ amplitude which includes gluonic radiative corrections \cite{LS92} as
well as the primordial transverse size of the bound state \cite{JK93}:
\begin{eqnarray}
  {\cal P}_{\pi}\left( x, b, Q, \mu _{\rm F}, \mu _{\rm R} \right)
= &&
 \exp \left[
            -   s\left( x, b, Q, \mu _{\rm F}, \mu _{\rm R} \right)
            -   s\left(\bar{x}, b, Q, \mu _{\rm F}, \mu _{\rm R}\right)
            - 2 \int_{\mu _{\rm F}}^{\mu _{\rm R}}
                \frac{{\rm d}\mu}{\mu}\,
                \gamma _{\rm q}\left(g(\mu )\right)
      \right]
\nonumber \\
&&
\times\; {\cal P}\left( x, b, \mu _{\rm F} \right) \; .
\label{eq:piamplbspace}
\end{eqnarray}
The pion distribution amplitude at the factorization point is
approximately given by
\begin{equation}
  {\cal P}\left( x, b, \mu _{\rm F} = C_{1}/b \right)
\simeq
  \phi _{\pi}\left( x, \mu _{\rm F} = C_{1}/b \right) \Sigma (x, b) \; ,
\label{eq:inpiampl}
\end{equation}
where $\Sigma (x,b)$ parametrizes the intrinsic transverse size of the
pion (see below). In the collinear approximation, one has
\begin{equation}
  \frac{f_{\pi}}{2\sqrt{2N_{\rm c}}} \,
  \phi _{\pi}\left( x, \mu _{\rm F}^{2} \right)
=
  \int_{}^{\mu _{\rm F}^{2}}
  \frac{{\rm d}{}^{2}{\bf k}_{\perp}}{16\pi ^{3}} \,
  \Psi _{\pi} ( x, {\bf k}_{\perp} ) \; ,
\label{eq:pida}
\end{equation}
where
$f_{\pi} = 130.7$~MeV and $N_{\rm c}=3$.
Integrating on both sides of this equation over $x$ normalizes
$\phi _{\pi}$
to unity, i.e.,
$
 \int_{0}^{1}{\rm d}x \, \phi _{\pi}\left( x, \mu _{\rm F}^{2} \right)
=
 1
$
because the rhs is fixed to
$
 \frac{f_{\pi}}{2\sqrt{2N_{\rm c}}} \,
$
by the leptonic decay $\pi \rightarrow \mu ^{+} \nu _{\mu}$ for any
factorization scale.

The Sudakov functions can be written in terms of the momentum-dependent
cusp anomalous dimension to read \cite{Kor89,Col89,KR92,GKKS97}
\begin{equation}
  s\left( \xi , b, Q, \mu _{\rm F}, \mu _{\rm R} \right)
=
  \frac{1}{2}
  \int_{\mu _{\rm F}=\frac{C_{1}}{b}}^{\mu _{\rm R}=C_{2}\xi Q}
  \frac{{\rm d}\mu}{\mu} \,
  \Gamma _{{\rm cusp}} \left( \gamma , g(\mu ) \right)
\; ,
\label{eq:sudfuncusp}
\end{equation}
where
$\gamma = \ln \left( \frac{C_{2}\xi Q}{\mu} \right)$
is the cusp angle, i.e., the emission angle of a soft gluon and the
bent quark line after injecting at the cusp point the external (large)
momentum by the off-mass-shell photon, and
\begin{equation}
  \Gamma _{{\rm cusp}} \left( \gamma , g(\mu ) \right)
=
    2 \ln \left( \frac{C_{2}\xi Q}{\mu} \right)
    A\left( g(\mu ) \right)
  + B\left( g(\mu ) \right) \; .
\label{eq:gammacusp}
\end{equation}
The functions $A$ and $B$ are defined by
\begin{eqnarray}
  A\left( g(\mu ) \right)
& = &
  \frac{1}{2}
  \left[
         2 \Gamma _{{\rm cusp}} \left( g(\mu ) \right)
       + \beta (g) \frac{\partial}{\partial g} {\cal K}(C_{1}, g(\mu ))
  \right]
\nonumber \\
& = &
       C_{F} \frac{\alpha _{\rm s}^{\rm an}(g(\mu ))}{\pi}
     + \frac{1}{2} K C_{\rm F}
       \left(
             \frac{\alpha _{\rm s}^{\rm an}(g(\mu ))}{\pi}
       \right)^{2} \; ,
\label{eq:funAnlo}
\end{eqnarray}
and
\begin{eqnarray}
  B\left( g(\mu ) \right)
& = &
  - \frac{1}{2}
    \left[
           {\cal K} \left(
                         C_{1}, g(\mu )
                   \right)
         + {\cal G} \left(
                         \xi, C_{2}, g(\mu )
                   \right)
    \right]
\nonumber \\
& = &
      \frac{2}{3} \frac{\alpha _{\rm s}^{\rm an}(g(\mu ))}{\pi}
      \ln \left(
      \frac{C_{1}^{2}}{C_{2}^{2}}\frac{{\rm e}^{2\gamma_{\rm E}-1}}{4}
          \right) \; ,
\label{eq:funBnlo}
\end{eqnarray}
respectively, and the $K$-factor in the ${\overline{\rm MS}}$ scheme
is given by the expression
\begin{equation}
  K
=
  \left( \frac{67}{18} - \frac{\pi ^{2}}{6} \right) C_{\rm A}
  - \frac{10}{9}n_{\rm f} T_{\rm F}
   + \beta _{0} \ln \left( C_{1} {\rm e}^{\gamma _{\rm E}}/2 \right)
\label{eq:Kfactor}
\end{equation}
where $C_{\rm A}=N_{\rm c}=3$, $n_{\rm f}=3$, $T_{\rm F}=1/2$,
$\gamma _{\rm E}$ being the Euler-Mascheroni constant.
The functions ${\cal K}$, ${\cal G}$ are calculable within
perturbative QCD \cite{CS81}. Both anomalous dimensions,
$
 \Gamma _{{\rm cusp}} (g(\mu ))
=
  C_{\rm F}\, \alpha _{\rm s}^{{\rm an}}(\mu ^{2})/\pi
$
and
$
  \gamma _{\rm q}\left( g(\mu ) \right)
=
 - \alpha _{\rm s}^{{\rm an}}(\mu ^{2})/\pi
$,
will be evaluated using the analytic model of \cite{SS97} in
next-to-leading logarithmic order (for more technical details, we refer
to \cite{SSK98}).

For simplicity, we follow \cite{JK93} and model the distribution
of primordial (intrinsic) transverse momentum in the pion wave function
in the form of a Gaussian normalized to unity:
\begin{equation}
  \Psi _{\pi}\left( x, {\bf k}_{\perp} \right)
=
  \frac{16 \pi ^{2} f_{\pi}}{2\sqrt{2N_{\rm c}}} \,
  \phi (x)\, \beta ^{2} g(x)
  \exp\left[ - g(x)\beta ^{2}{\bf k}_{\perp}^{2} \right] \; ,
\label{eq:piwfjk}
\end{equation}
where $g(x)=1/x\bar{x}$ and the quark masses are neglected. For
$\phi _{\rm as}(x)=6x\bar{x}$,
one has
$\beta ^{2}=0.883~[{\rm GeV}^{-2}]$
which corresponds to
$<{\bf k}_{\perp}^{2}>^{1/2}=350$~MeV.

\begin{figure}
\centerline{\epsfig{file=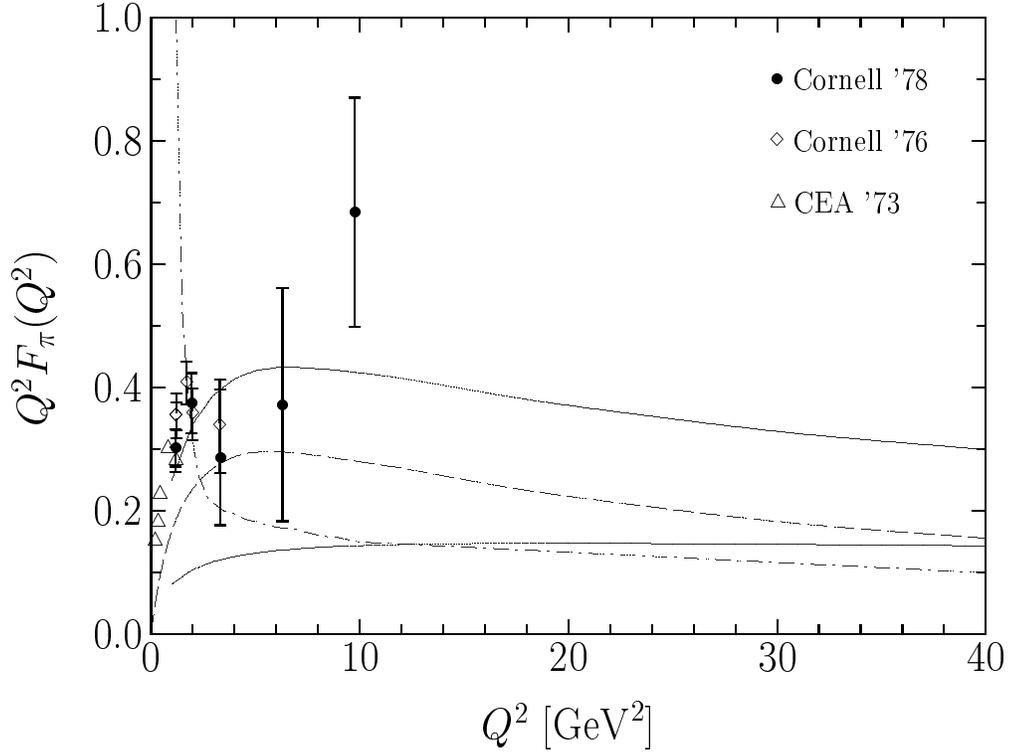,height=10.0cm,width=14.0cm,silent=}}
\vspace{0.5cm}
\caption[fig:pionffall]
        {\tenrm Spacelike pion form factor calculated with
        $\phi _{\rm as}$ within different factorization schemes, except
        for the long-dashed line which shows the soft contribution
        \cite{JKR96}, in comparison with the existing experimental
        data \cite{Bro73,Beb76}.
        The lower solid line shows the IR-enhanced result obtained
        in our analysis, whereas the upper one stands for the
        sum of the long-dashed and the lower solid line.
        The dot-dashed line reproduces the calculation of \cite{TL98},
        which does not include an intrinsic $k_{\perp}$-dependence.
\label{fig:piffirf}}
\end{figure}
%

Before we proceed with the presentation of our results, exposed in
Fig. \ref{fig:piffirf}, let us at this point interject some comments
regarding the role of the scales entering the calculation of the pion
form factor. Whenever
$\xi < \frac{C_{1}}{C_{2}}\frac{1}{bQ}$,
all Sudakov exponential factors are set equal to unity \cite{LS92}.
For all values of $b$, there is a hierarchy of scales according to
$
 \Lambda _{\rm QCD}^{\rm an} \ll \mu _{\rm F} = C_{1}/b
 \leq \mu _{\rm R} = C_{2}\xi Q \leq Q
$.
The limit $\mu _{\rm R} \simeq \mu _{\rm F}$ can be interpreted as the
minimum virtuality scale of exchanged quanta (or equivalently as the
maximum transverse separation) in $T_{\rm H}$ below which propagators
cannot be treated within perturbation theory and are therefore absorbed
into $\phi _{\pi}$.
In the present analysis we use
$\mu _{\rm F} = C_{1}/b$, $\mu _{\rm R} = C_{2}\xi Q$
with
$\xi = x, \bar{x}, y, \bar{y}$, and
$C_{1} = \exp\left[ -\frac{1}{2}(2\gamma _{\rm E} - 1) \right]$,
$C_{2} = 1/\sqrt{2}$, the latter corresponding to the value
$C_{2}^{CS} = 1$ (cf. \cite{CS81}).
For this choice of the scheme constants, the logarithmic term in the $K$
factor is eliminated.
Note that for technical reasons, the argument of $\alpha _{\rm s}$ in
$T_{\rm H}$ is taken to be $\mu _{\rm R} = C_{2}\sqrt{\xi\bar \xi}Q$,
where $\xi$ is either $x$ or $y$.

\begin{figure}
\centerline{\epsfig{file=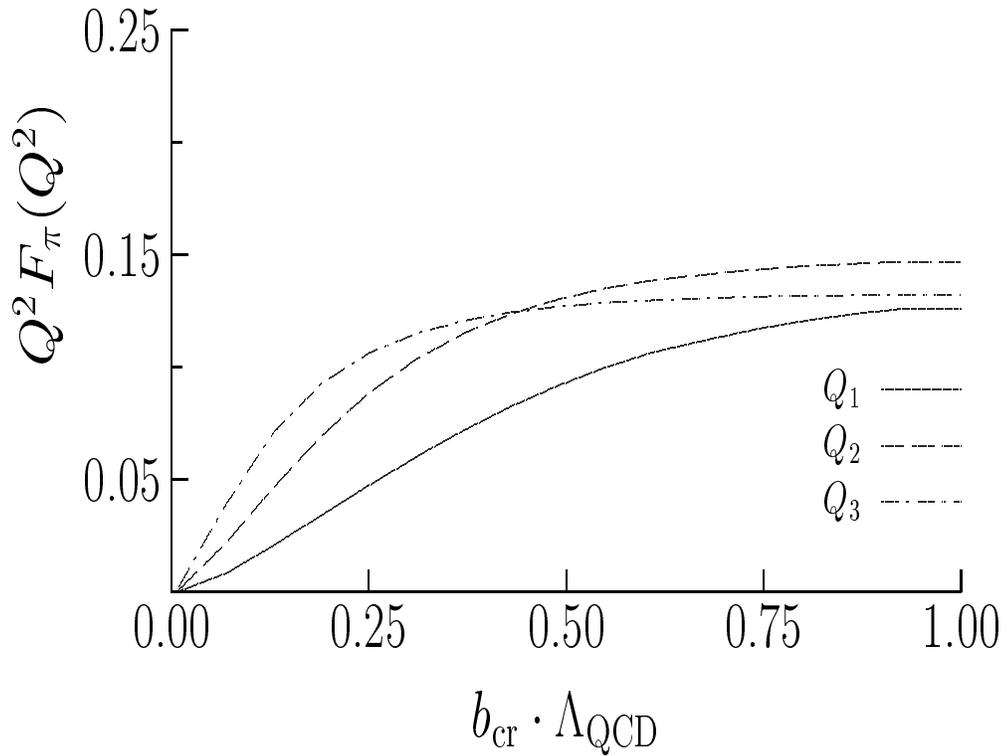,height=10.0cm,width=14.0cm,silent=}}
\vspace{0.5cm}
\caption[fig:pionffall]
        {\tenrm Dependence on the (critical) transverse distance $b$ of
        the hard contribution to the scaled pion form factor, calculated
        with $\phi _{\rm as}$ within our IR-finite factorization scheme,
        for three different values of the momentum transfer:
        $Q_{1}=2$~GeV (solid line), $Q_{2}=5$~GeV (dashed line), and
        $Q_{3}=10$~GeV (dashed-dotted line).
\label{fig:satold}}
\end{figure}

In order to show how the contribution to the pion form factor is
accumulated in $b$ space, we show in Fig. \ref{fig:satold} the
dependence of the scaled pion form factor against
$b\,\Lambda _{\rm QCD}$. One observes a fast rise of the displayed
curves as $Q$ increases.
Indeed, already for the smallest value shown, $Q=2$~GeV, the form
factor accumulates half of the whole contribution in the region
$b\simeq 0.5/\Lambda _{\rm QCD}$. For still larger $Q$ values, the form
factor levels off already around $b=0.3/\Lambda _{\rm QCD}$ for
$Q=5$~GeV, and $b=0.25/\Lambda _{\rm QCD}$ for $Q=10$~GeV. This behavior
of the curves uncovers how the IR stability and self-consistency of the
perturbative treatment in the present scheme is improved, as compared to
previous, conventional, approaches \cite{LS92,JK93,TL98}.

%

\begin{figure}
%
\centerline{\epsfig{file=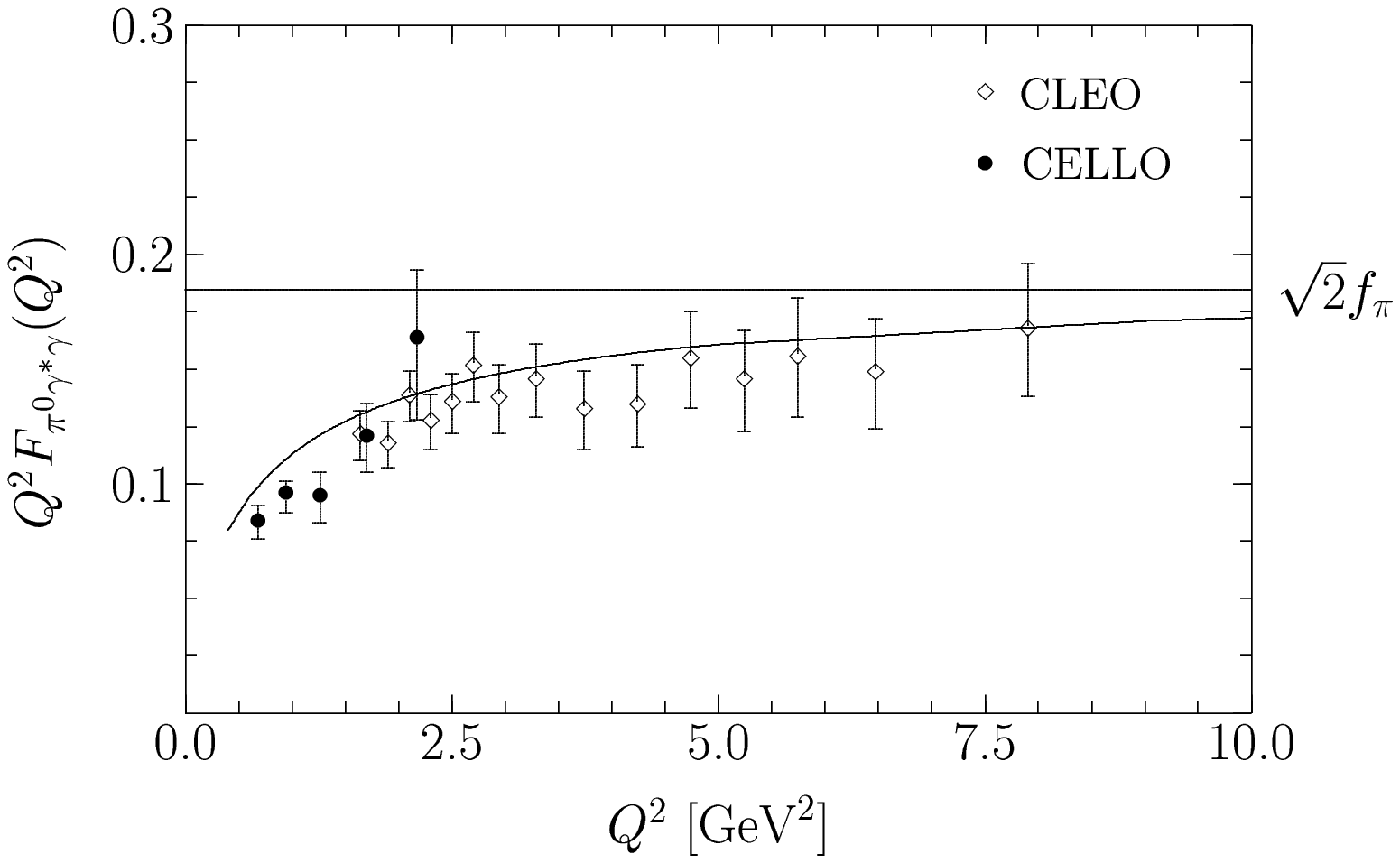,height=10.0cm,width=14.0cm,silent=}}
\vspace{0.5cm}
\caption[fig:piongamma]
        {\tenrm Pion-photon transition form factor calculated with
        $\phi _{\rm as}$ and IR enhancement (lower solid line).
        The other solid line shows the asymptotic behavior.
        The data are taken from \cite{CLEO98,CELLO91}.
\label{fig:pigairf}}
\end{figure}

A similar expression to Eq.~(\ref{eq:piffbspace}) holds also for
$F_{\pi ^{0}\gamma ^{*}\gamma}$,
the main difference being that the latter contains only one pion wave
function, and furthermore the associated short-distance part,
$T_{\rm H}$, does not depend on $\alpha _{\rm s}$ in LO.
The only dependence on the strong coupling constant at leading
order enters through the anomalous dimensions of the cusp and the quark
wave function. The result of this calculation is displayed in
Fig. \ref{fig:pigairf}. Notice that in this case, we use
$\mu _{\rm R}=C_{2} \xi Q$, where $\xi = x$ or $\xi =\bar x$.

%

In summary, we have shown that modifying $\alpha _{\rm s}$ in
the IR region by a nonperturbative power correction, which removes the
unphysical Landau singularity, may play a crucial role in the
{\it practical} calculation of exclusive processes because it improves
the IR stability of computed observables based on perturbation
expansions without introducing external parameters to ``freeze'' the
running strong coupling. Furthermore, we have given quantitative
evidence that in this way it is possible to get an enhanced hard
contribution to $F_{\pi}(Q^{2})$, relying exclusively on the asymptotic
form of the pion distribution amplitude, so that, though this
contribution comprises Sudakov corrections and a primordial
$k_{\perp}$-dependence, it is not suppressed.
Together with the soft part of $F_{\pi}(Q^{2})$, this contribution can
account for the trend of the existing (admittedly low-accuracy) data
without employing endpoint-concentrated pion distribution amplitudes.
The same treatment yields for $F_{\pi ^{0}\gamma ^{*}\gamma}$ a
theoretical prediction which is in good agreement with the data.

\bigskip
\ac
It is a pleasure to thank Peter Kroll for useful discussions and
remarks.
The work of H.-Ch.K. was supported in part by the Research Institute for
Basic Sciences, Pusan National University under Grant RIBS-PNU-98-203.
\newpage   

\newpage   
\end{document}